\documentclass[twocolumn,showpacs,preprintnumbers,amsmath,amssymb,aps,pra]{revtex4-1}
\usepackage{lipsum}
\usepackage{graphicx}
\usepackage{epstopdf}
\usepackage{dcolumn}
\usepackage{bm}
\usepackage{float}
\usepackage{amsmath}

\begin{document}     
      

\title{Two quantum particles trapped in three dimensions harmonic oscillator and  interacting via finite range soft-core interaction}

\author{M. A. Shahzad}
 \affiliation{ Department of Physics, Hazara University, Pakistan.}
\email{Email:muhammad.shahzad@unicam.it}

\date{\today}

\begin{abstract}
We study the exactly solvable quantum system of two  particles  confined
in a three-dimensional harmonic trap and interacting via finite-range soft-core interaction by means of the separation of variables and ansatz
method. Supposing the solution in the form of  ansatz $\Psi(r)=r^{a}e^{-\lambda r^2} \psi(r)$ we transform the time independent Schr\"{o}dinger equation into Kummer's differential equation whose solution are given in the form of confluent hypergeometric function. We also discuss that in the absence of central force potential, the quantum system map into the problem  of two quantum particle trapped in one-dimension harmonic oscillator and interacting through finite distance soft-core potential. In such special case the eigen value equation  become the Weber's differential equation  and its solution are also given in the form of  confluent hypergeometric function.

\end{abstract}


\maketitle
\section{Introduction}-

Exactly solvable potentials play a very important role in various fields of Physics. These potentials
served as useful tools in modeling realistic physical problems, and offered an interesting
field of investigation in different fields of applications in Physics. \\
\indent The quantum system of  particles with
short range interactions in a harmonic oscillator has been
well studied. Identical particles in such a trap
are referred to as harmonium \cite{J}, which has been
studied as an exactly solvable model for the artificial helium
atom, where the interaction of the electrons with the nucleus has been replaced by a harmonic confinement. 
Cold atomic gases offer a promising route towards ultimate
brightness conditions. In particular Rydberg crystals
are very close to the quantum plasma regime, and
extraction of the electrons could be done with a minimum
amount of heating, such that the electrons could
be trapped in Paul traps and
Ponderomotive traps.\\
\indent Exactly solvable quantum  model \cite{bloch1} of two ultra-cold bosons confined in a harmonic trap and interacting
via contact forces play important role to understand strongly correlated many body systems. Particularly, exact solutions of the two particle  model were essential for understanding of the
Tonks-Girardeau limit of infinite repulsions between particles \cite{bloch2}. Recently, P. Ko\'{s}cik and  T. Sowi\'{n}ski  studies the exactly solvable model of two indistinguishable quantum particles (bosons or fermions) confined
in a one-dimensional harmonic trap and interacting via finite-range soft-core interaction \cite{bloch3}. They shown that independently on the
potential range, in the strong interaction limit bosonic and fermionic solutions become degenerate. \\
\indent Here, we present a generalization of the one-dimensional  P. Ko\'{s}cik and  T. Sowi\'{n}ski model of two quantum
particles (bosons as well as fermions) interacting via the force of a finite range  to the case of two particles in three-dimensions harmonic potential. We present analytically the  solutions of the Schr\"{o}dinger equation  for the two quantum particle interacting via soft-core interaction and trapped in three-dimension harmonic oscillator. We map the three-dimensions eigenvalue  equation into the Kummer's differential equation and presents its  solution in term of  confluent hypergeometric function.

\indent

\section{The Eigenvalue problem}

Consider a quantum system consists of two identical particles of mass $m$ interacting through soft-core potential and trapped in harmonic potential of frequency $\omega$. The Schr\"{o}dinger equation for the system is 

\begin{align}
 &-{\hbar^2\over 2m}\Big[{\partial^2 \Psi\over\partial x_1^2}+{\partial^2 \Psi\over\partial x_2^2}\Big]+\Big[{m\omega^2\over 2}\big(x_1^2+x_2^2\big)+V(x_1-X_2)\Big]\Psi\nonumber\\
&\qquad=E\Psi
\end{align}

where the interaction potential $V(x)$ is 

\[
 V(x) = 
  \begin{cases} 
   V & \text{if } |x| < a \\
   0      & \text{if }  |x|\geq a
  \end{cases}
\]

Using center of mass coordinate,
\begin{equation}
R={x_1+x_2\over 2},\qquad r=x_2-x_1
\end{equation}
With the introduction of these new variables, the Schr\"{o}dinger equation can written as a sum of two independent single particle equation with Hamiltonian $H=H_R+H_r$,
\begin{eqnarray}\label{eq:Har}
-{\hbar^2\over 4m}{d^2 \Psi\over dR^2}+m\omega^2R^2\Psi=E\Psi\\ 
-{\hbar^2\over m}{d^2 \Psi\over dr^2}+V(r)\Psi=E\Psi
\end{eqnarray}
Equation (3) have the form of  schr\"{o}dinger equation with harmonic potential, whose solution can be easily find. The second equation has an additional term due to soft-core interaction $V(x)$ and can be rewritten using natural units of the an external harmonic oscillator, that is 
\begin{equation}
\Big[-{d^2\over dr^2}+{1\over 4}r^2+V(r)\Big]\Psi=E\Psi
\end{equation}
In three dimensions, the above time independent Schr\"{o}dinger equation takes the form
\begin{align}
&\Big[  -{d^2\Psi\over dx^2 } -{d^2\Psi\over dy^2 } -{d^2\Psi\over dz^2 }\Big] \nonumber\\
&+\Big\{{1\over 4}(x^2+y^2+z^2) +V(x,y,z) \Big\}            \Psi(x,y,z)=E\Psi(x,y,z)
\end{align}
In spherical coordinates, the time  independent Schr\"{o}dinger equation is given by
\begin{align}
&-\Big[{1\over r^2}   {\partial \over \partial r}\Big( r^2{\partial \Psi\over \partial r}    \Big)  +{1\over r^2\sin\theta} {\partial\over \partial \theta}    \Big(\sin\theta {\partial\Psi\over\partial \theta}  \Big) \Big.\nonumber\\
 &+{1\over r^2\sin^2\theta} \Big({\partial^2\over \partial\phi^2}    \Big]+{1\over 4}r^2\Psi+V(r)\Psi=E\Psi
\end{align}
Using method of separations of variables, we seek a solution of the form $\Psi(r,\theta,\phi)=R(r)Y(\theta,\Phi)$. Substituting in the above equation and separating yields the radial and angular equations;
\begin{eqnarray*}
{d\over dr}\Big(r^2{dR\over dr} \Big)-r^2\Big[{r^2\over 4}+V(r)-E  \Big]R=l(l+1)R\\
{1\over Y}\Big\{ {1\over \sin\theta} {\partial\over \partial \theta}    \Big( \sin\theta{\partial Y\over\partial \theta}  \Big)  + {1\over \sin^2\theta} \Big({\partial^2 Y\over \partial\phi^2}    \Big)     \Big\}=-l(l+1)
\end{eqnarray*}
The solution of angular equation is called spherical harmonics. Using $u(r)=rR(r)$, equation (8) can be written as
\begin{equation}
-{d^2 u\over d r^2}+\Big[ V(r) +{1\over 4} r^2+{l(l+1)\over r^2}   \Big]u=Eu
\end{equation}
where the effective potential
\begin{equation}
V_{eff}= V(r) +{1\over 4} r^2+{l(l+1)\over r^2}   
\end{equation}
As a special care  with $l=0$, equation (8) has the form of Weber differtail equation
\begin{equation}
  \Big[-{d^2\over d r^2}+{1\over 4}r^2\Big]u=- \Big[\eta+{1\over 2}\Big]u
\end{equation}
with $\eta=-E-V-1/2$ for $|r|<a$ and $\eta=-E-1/2$ for $|r|\geq a$. For  non-interacting partiales (V=0), we get a solution for harmonic oscillator, that is
\begin{equation}
u_n(r)=N_n e^{-r^2/4} H_n(r/\sqrt{2}).
\end{equation}
where $H_n(x)$ is the Hermit polynomial.\\
For $V\neq 0$, the solution of Weber differtial equation can be written in term of the confluent hypergeometric function $_1F_1$,
\begin{equation}
u^{(+)}(r)=e^{-r^2/4}  {_1F_1}\Big[ {{\eta+1}\over 2} ; {1\over 2}; {r^2\over 2}  \Big]
\end{equation}

\begin{equation}
u^{(-)}(r)=re^{-r^2/4}  {_1F_1}\Big[ {{\eta+2}\over 2} ; {3\over 2}; {r^2\over 2}  \Big]
\end{equation}
These function are divergent in the infinity, $r\rightarrow \pm\infty$, and hence the appropriate solutions exists only in the region $|r|< a$.\\

With Coulomb interaction of  the form $V=1/r$, the effective potential is 
\begin{equation}
V_{eff}= {1\over r} +{1\over 4} r^2+{l(l+1)\over r^2}   
\end{equation}
The corresponding Schro\"{o}dinger equation can be written as
\begin{equation}
-{d^2 u\over d r^2}+\Big[ {1\over r} +{1\over 4} r^2+{l(l+1)\over r^2}   \Big]u=Eu
\end{equation}
The properties of Eq.(15) with effective potential given in Eq.(14) were discussed in detail in the literature \cite{C1,C2,C3,C4} and the corresponding quantum system is referred to as harmonium. The  Schro\"{o}dinger equation Eq.(15) is not analytically solvable, but rather the problem is quasi-exactly solvable. In such  quasi-exactly solvable limit the wave functions can be written as

\begin{equation}
u(r)=r^{l+1}e^{-r^2/4}P(r)
\end{equation}
where 

\begin{equation}
P(r)=\sum_{i=0}^{p}a_ir^i, \qquad \lim_{r\rightarrow\infty}u(r)=0. 
\end{equation}
is a polynomial of finite degree. With this Ansatz, Eq. (15)
gives a recurrence relation for the coefficients $a_i$, which
can be solved when $i$ is given.

Now, Eq.(8) can be rewritten as
\begin{equation}
{d^2 u\over d r^2}+\Big[ \chi -{1\over 4} r^2-{l(l+1)\over r^2}   \Big]u=0
\end{equation}
where $\chi=E-V$ for $|r|<a$ and $\chi=E$ for $|r|\geq a$. Substituting
\begin{equation}
u(r)=r^{l+1}e^{-r^2/2}F(r)
\end{equation}
in equation (18), we find that $F(x)$ satisfies Kummer's differential equation
\begin{equation}
x{{d^2F\over {dx^2}}}+(b-x){{dF\over {dx}}}-aF=0
\end{equation}
where $x=r^2$, $a=2l+3-\chi$, and $b=l+3/2$. The solution of Kummer's differential equation are confluent hypergeometric function \cite{F1,F2}
\begin{align}
M(a,b,x)&=1+{a\over b}x+{a(a+1)\over b(b+1)}{x^2\over 2!}\nonumber\\
&+ {a(a+1)(a+2)\over b(b+1)(b+2)}{x^3\over 3!}+....=_1F_1(a;b;x)
\end{align}
with a polynomial structure of when $a=-n_r$, with $n_r=0,1,2,....$. The quantum number $n_r$ is called the radial quantum number. From $a=2l+3-\chi$, we have
\begin{equation}
\chi=4n_r+2l+3=n
\end{equation}
The positive integer $n$ is called the principle quantum number. 
The solution of  the above equation for angular momentum $l$ can be written in term of confluent hypergeometric function, 
\begin{equation}
u=r^{l+1}e^{-r^2/2} M\Big({{2l+3-n}\over 4},l+{3\over 2},r^2\Big)
\end{equation}
Figure (1) shows the plot of the radial wave function $u(r)$ as a function of $r$ for differential value of $n$ and $l$. It should be noted that the radial wave functions with lowest value of $l$ for a given value of $n$, has no nodes for $r>0$. The outermost maximum of the each wave function is seen to occur at decreasing distances from the origin as $n$ increases.

\begin{figure}[t]
\centering
\includegraphics[scale=0.71]{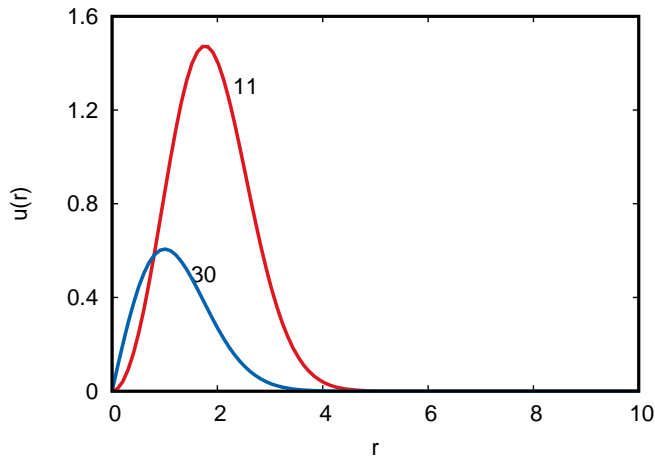}\\
\caption{ Radial wave function $u(r)$ (Eq.(23)) as a function of $r$ for  $n=1,3$ and $l=0,1$.\label{LBP}}
\label{fig:digraph}
\end{figure}

\section{Conclusion} 
We studied  the exactly solvable quantum system of two  particles  trapped
in a three-dimensional harmonic oscillator and interacting via finite-range soft-core interaction using method of  separation of variables. The time independent Schr\"{o}dinger equation are mapped into the Kummer's differential equation and its solution are presented  in the form of confluent hypergeometric function. With in the limit of special case $l=0$, we obtained a model  of two quantum particle trapped in one-dimension harmonic oscillator and interacting through finite distance soft-core potential.  The eigen value equation  become the Weber's differential equation which was basically used to solve Laplace equation expressed in parabolic coordinates, and its solution are given form of  confluent hypergeometric function.  The results will enabled us to investigate  different properties of the system in a whole range of parameters between limiting cases of Busch \textit{et al.}  and hard-core models. The model can be used to understand ground states of many-body system in three-dimension.

\indent

\end{document}